\PassOptionsToPackage{dvipsnames}{xcolor}
\documentclass[RNAAS]{aastex631}

\usepackage{xspace} 
\usepackage{natbib}

\newcommand{\snap}{(a)\xspace}
\newcommand{\survival}{(b)\xspace}
\newcommand{\mass}{(c)\xspace}

\usepackage{hyperref}

\begin{document}

\title{Survival of higher overdensity cold gas in a turbulent, multiphase medium}

\author[0009-0005-4937-100X]{Ashwin Vergis George}
\affiliation{University of Potsdam, Potsdam, Germany}
\email{ashwin.george@uni-potsdam.de}

\author[0000-0001-6171-6924]{Hitesh Kishore Das}
\affiliation{Max Planck Institute for Astrophysics,
Garching, Germany}

\author[0000-0003-2491-060X]{Max Gronke}
\affiliation{Astronomisches Rechen-Institut, Zentrum für Astronomie, Heidelberg University, Heidelberg, Germany}
\affiliation{Max Planck Institute for Astrophysics, 
Garching, Germany}

\begin{abstract}

Cold gas clouds embedded in a hot, turbulent medium are typically short-lived due to disruptive hydrodynamic instabilities. However, radiative cooling might allow such clouds to survive and grow. We present 3D \texttt{Athena++} simulations of clouds with a density contrast of $\chi = 1000$, exploring turbulent Mach numbers $\mathcal{M}\in (0.25, 0.75)$ and cloud radii chosen to span cooling-to-crushing ratios $\alpha \in [0.001, 10]$. We find a shift in the survival boundary, with cloud survival occurring only when the cooling-to-cloud-crushing ratio ($t_{\text{cool,mix}} / t_{\text{cc}}$) $\lesssim 0.01$, which is lower than the expected boundary of $\sim 1$. This result shows that it is more difficult for higher over-density cold clouds to survive in a turbulent, hot medium, and suggests another `survival criterion'.
\end{abstract}


\section{Introduction} \label{sec:intro}

The multiphase nature of astrophysical media, such as the intra-cluster medium, is well established. In the ICM, AGN feedback drives a baryon cycle that regulates both thermal balance and the energy budget  \citep{donahue2022}, shaping the evolution of these systems. The ICM is not only multiphase but also highly turbulent, which plays a key role in cold-phase growth and the mass budget.

Previously, \citet{gronketurb2022} found that cold gas clouds can grow if the cooling timescale $t_{\text{cool,mix}} $(at $\sqrt{T_{\text{hot}} T_{\text{cold}}} $ and $\sqrt{\rho_{\text{hot}} \rho_{\text{cold}}}$), is shorter than the cloud-crushing time, $t_{\rm cc} = \chi^{1/2} R_{\rm cl}/v_{\rm turb}$, where $v_{\rm turb}$ is the mass-weighted RMS turbulent velocity and $\chi \equiv \rho_{\rm cold}/\rho_{hot}$. Similar studies which have investigated this survival of cold gas have focused on the CGM (temperatures $\sim10^4$--$10^6$~K and  $\chi$ $\sim 100$). However, analogous studies in the hotter ICM, with $\chi \sim 1000$, are still comparatively sparse.

In this study, we aim to extend and test the evolution of these cold clouds in the ICM.
\begin{figure}[htbp]
    \centering
    \includegraphics[width=\linewidth]{combined_figure.pdf}
    \caption{ \textbf{a)} Density projections at 3 different times. Top: $\alpha=1$, cold cloud is destroyed; Bottom: $\alpha=0.01$ cold cloud grows.\textbf{ b)} Cold cloud gas mass normalised with initial mass, $m_{\text{initial}}$ vs time normalised by $t_{\text{eddy}}$. Solid lines:$128^3$ cells; dashed lines: $256^3$ cells at $\mathcal{M} \sim 0.5$. $\textbf{c)}$ Plot of $\mathcal{M}$ vs $\alpha$, points at $\mathcal{M}\sim0.5$ were offset for visual clarity. The grey line shows the established survival boundary for clouds for $\chi$ of 100 from \citep{gronketurb2022}, the black line represents the observed survival boundary for $\chi=1000$.    }
    \label{fig:image}
\end{figure}

\section{Methods} \label{sec:methods}
The simulation setup follows \citet{das2024}, using \texttt{Athena++} to study cold gas clouds a temperature $T_{\rm cold} = 4 \times 10^4$K and density $\rho_{\rm cold} = \chi \rho_{\rm hot}$ embedded in a turbulent, hot medium at temperature $T_{\rm hot}=4\times 10^7$K and density $\rho_{\rm hot}$. $L_{\text{box}}/r_{\rm cl}$ is set as $40$.
Hereafter, we define $\alpha$ as the ratio $ \alpha \equiv {t_{\text{cool,mix}}} / {t_{\text{cc}}}$. 

\section{Results and Discussion} \label{sec:Results} 
We perform 33 simulations with $\mathcal{M}\equiv v_{\rm turb}/c_{\rm s,hot} \in (0.25, 0.75)$ and $\alpha \in (0.001,10)$. We repeat simulations with different random seed values at $\mathcal{M} = 0.5$ to study the effect of the stochasticity of turbulence. 
Since the turbulent fields can be rescaled to a required box size, we rescale and restart the same snapshot at the end of the turbulent driving phase for each turbulent Mach number.


Fig.~\ref{fig:image}\snap shows density projections for two cold cloud sizes, at the different normalised times $t/t_{\text{eddy}}$ in columns (at a resolution of $256^3$ cells). The top row depicts a cold cloud with $\alpha = 1$, which is destroyed as turbulent mixing dominates over cooling. Whereas the bottom row shows a cold cloud with $\alpha = 0.01$, whose mass grows. Here the mixed gas cooling dominates and the cold gas mass increases with time. This comparison highlights the morphological evolution of the cold clouds.

Fig.~\ref{fig:image}\survival shows the temporal evolution of the cold gas mass for the cases with a turbulent Mach number, $\mathcal{M}=0.5$. We define cold gas mass as the total mass of gas in the box with temperature below $2T_{\rm floor}=8\times 10^4$K. The colour of the lines represents different values of $\alpha$, with solid and dashed lines representing our fiducial ($128^3$ cells) and high-resolution ($256^3$ cells) simulations. We find that the general trend for cloud survival is consistent across resolutions and turbulent driving random seeds. This figure also shows the existence of a critical $\alpha$ as a survival criterion, below which the cold gas cloud survives in a turbulent hot medium. 

Fig.~\ref{fig:image}\mass summarises the survival criterion across all runs by displaying Mach number versus the ratio $t_{\text{cool,mix}} / t_{\text{cc}}$, with the colour of the point denoting the final cold mass ratio. At $\mathcal{M} \approx 0.5$, we show simulations with different turbulent driving random seeds, as shown in Fig.~\ref{fig:image}\survival. 
We find that the mass of the cold cloud only increases for $\alpha \lesssim 0.01$. 
The light grey line indicates the expected survival criterion. This line is the $\chi=1000$ version of the survival criterion found with $\chi = 100$ simulations in \citep{gronketurb2022}. We find that a different survival criterion, shifted by $\sim 1$ dex, shown as the dark grey line, divides the survival and destruction regimes for our simulations. However, further exploration is needed to characterise the exact extent of this shift. Both curves exhibit a downward slope due to a higher probability of destruction at higher Mach numbers \citep[see discussion in][]{gronketurb2022}. 

\section{Conclusion} \label{sec:conclusion}

In this study, we observe a systematic trend of survival and destruction across various $\mathcal{M}$ and resolutions and a 1 dex shift from the expected survival criterion. Hence, $\chi=1000$ clouds need to cool more rapidly, in comparison to $\chi=100$ clouds, to survive.

Some have claimed similar deviations from the $t_{\rm cool,mix}/t_{\rm cc}$ criterion for $\chi\gtrsim 1000$ \citep{Sparre2019TheWind,abruzzo2022taminginteractions} whereas others do not \citep{gronke2018thewind,farber2021thewinds}. `Falling clouds' simulations suggest a criterion $\sim t_{\rm grow}/t_{\rm cc}$ (where $t_{\rm grow}\equiv m/\dot m$) implying a lower survivability for higher $\chi$ systems \citep{tan2023infalling}. These authors justify such lower survivability in comparison to the `cloud-crushing' studies with the inability to collect mixed gas in lower shear regions (the tail of the cloud) and the overall evolution of the system: while for a ram pressure accelerated cloud the shear drops, i.e., it becomes easier for the cold gas to survive as time progresses, this is not the case for an infalling cloud. A similar argument can be brought up in the turbulent boxes studied here, where the flow remains turbulent at all times, so no quiescent region (or time) forms where the shear could drop. Consequently, the lack of a low shear region reduces the survivability of these cold clouds as compared to ram-pressure accelerated clouds. However, further numerical studies are needed to quantify this and come to a firm conclusion.

\section{Acknowledgments}
MG thanks the Max Planck Society for support through the MPRG, and the EU for support through ERC-2024-STG 101165038 (ReMMU). HD thanks the staff and colleagues at MPA and also IMPRS for their valuable support. AG thanks the co-authors for their mentorship and guidance.
\facility{FREYA at MPCDF.}
\bibliography{references, sample631}{}

\begin{thebibliography}{}
\expandafter\ifx\csname natexlab\endcsname\relax\def\natexlab#1{#1}\fi
\providecommand{\url}[1]{\href{#1}{#1}}
\providecommand{\dodoi}[1]{doi:~\href{http://doi.org/#1}{\nolinkurl{#1}}}
\providecommand{\doeprint}[1]{\href{http://ascl.net/#1}{\nolinkurl{http://ascl.net/#1}}}
\providecommand{\doarXiv}[1]{\href{https://arxiv.org/abs/#1}{\nolinkurl{https://arxiv.org/abs/#1}}}

\bibitem[{{Abruzzo} {et~al.}(2024){Abruzzo}, {Fielding}, \& {Bryan}}]{abruzzo2022taminginteractions}
{Abruzzo}, M.~W., {Fielding}, D.~B., \& {Bryan}, G.~L. 2024, \apj, 966, 181, \dodoi{10.3847/1538-4357/ad1e51}

\bibitem[{{Begelman} \& {Fabian}(1990)}]{begelman1990}
{Begelman}, M.~C., \& {Fabian}, A.~C. 1990, \mnras, 244, 26P

\bibitem[{{Binney} \& {Tabor}(1995)}]{AGNBinney1995MNRAS.276..663B}
{Binney}, J., \& {Tabor}, G. 1995, \mnras, 276, 663, \dodoi{10.1093/mnras/276.2.663}

\bibitem[{Brandenburg \& Nordlund(2011)}]{Brandenburg2011AstrophysicalModeling}
Brandenburg, A., \& Nordlund, A. 2011, Reports on Progress in Physics, 74, 046901, \dodoi{10.1088/0034-4885/74/4/046901}

\bibitem[{Burkhart {et~al.}(2020)Burkhart, Appel, Bialy, Cho, Christensen, Collins, Federrath, Fielding, Finkbeiner, Hill, Ib{\'{a}}{\~{n}}ez-Mej{\'{i}}a, Krumholz, Lazarian, Li, Mocz, Mac~Low, Naiman, Portillo, Shane, Slepian, \& Yuan}]{Burkhart2020TheCATS}
Burkhart, B., Appel, S.~M., Bialy, S., {et~al.} 2020, The Astrophysical Journal, 905, 14, \dodoi{10.3847/1538-4357/abc484}

\bibitem[{{Ciotti} \& {Ostriker}(2001)}]{AGNCiotti2001ApJ...551..131C}
{Ciotti}, L., \& {Ostriker}, J.~P. 2001, \apj, 551, 131, \dodoi{10.1086/320053}

\bibitem[{{Das} \& {Gronke}(2024)}]{das2024}
{Das}, H.~K., \& {Gronke}, M. 2024, \mnras, 527, 991, \dodoi{10.1093/mnras/stad3125}

\bibitem[{{Donahue} \& {Voit}(2022)}]{donahue2022}
{Donahue}, M., \& {Voit}, G.~M. 2022, \physrep, 973, 1, \dodoi{10.1016/j.physrep.2022.04.005}

\bibitem[{Elmegreen \& Scalo(2004)}]{Elmegreen2004InterstellarProcesses}
Elmegreen, B.~G., \& Scalo, J. 2004, Annual Review of Astronomy and Astrophysics, 42, 211, \dodoi{10.1146/annurev.astro.41.011802.094859}

\bibitem[{Eswaran \& Pope(1988)}]{eswaran1988257}
Eswaran, V., \& Pope, S. 1988, Computers \& Fluids, 16, 257, \dodoi{https://doi.org/10.1016/0045-7930(88)90013-8}

\bibitem[{Falceta-Gon{\c{c}}alves {et~al.}(2014)Falceta-Gon{\c{c}}alves, Kowal, Falgarone, \& Chian}]{FalcetaGoncalves2014}
Falceta-Gon{\c{c}}alves, D., Kowal, G., Falgarone, E., \& Chian, A. C.-L. 2014, Nonlinear Processes in Geophysics, 21, 587, \dodoi{10.5194/npg-21-587-2014}

\bibitem[{Farber \& Gronke(2021)}]{farber2021thewinds}
Farber, R.~J., \& Gronke, M. 2021, arXiv e-prints, 17, 1.
\newblock \url{http://arxiv.org/abs/2107.07991}

\bibitem[{{Faucher-Gigu{\`e}re} \& {Oh}(2023)}]{faucher2023}
{Faucher-Gigu{\`e}re}, C.-A., \& {Oh}, S.~P. 2023, \araa, 61, 131, \dodoi{10.1146/annurev-astro-052920-125203}

\bibitem[{Federrath(2013)}]{Federrath2013OnTurbulence}
Federrath, C. 2013, Monthly Notices of the Royal Astronomical Society, 436, 1245, \dodoi{10.1093/mnras/stt1644}

\bibitem[{{Field} {et~al.}(1969){Field}, {Goldsmith}, \& {Habing}}]{TwoPhase1969ApJ...155L.149F}
{Field}, G.~B., {Goldsmith}, D.~W., \& {Habing}, H.~J. 1969, \apjl, 155, L149, \dodoi{10.1086/180324}

\bibitem[{{Gronke} {et~al.}(2022){Gronke}, {Oh}, {Ji}, \& {Norman}}]{gronketurb2022}
{Gronke}, M., {Oh}, S.~P., {Ji}, S., \& {Norman}, C. 2022, \mnras, 511, 859, \dodoi{10.1093/mnras/stab3351}

\bibitem[{Gronke \& Peng~Oh(2018)}]{gronke2018thewind}
Gronke, M., \& Peng~Oh, S. 2018, Monthly Notices of the Royal Astronomical Society: Letters, 480, L111, \dodoi{10.1093/mnrasl/sly131}

\bibitem[{Harris {et~al.}(2020)Harris, Millman, van~der Walt, Gommers, Virtanen, Cournapeau, Wieser, Taylor, Berg, Smith, Kern, Picus, Hoyer, van Kerkwijk, Brett, Haldane, del R{\'{i}}o, Wiebe, Peterson, G{\'{e}}rard-Marchant, Sheppard, Reddy, Weckesser, Abbasi, Gohlke, \& Oliphant}]{harris2020arraynumpy}
Harris, C.~R., Millman, K.~J., van~der Walt, S.~J., {et~al.} 2020, Nature, 585, 357, \dodoi{10.1038/s41586-020-2649-2}

\bibitem[{Hunter(2007)}]{hunter2007matplotlib:environment}
Hunter, J.~D. 2007, Computing in Science {\&} Engineering, 9, 90, \dodoi{10.1109/MCSE.2007.55}

\bibitem[{Kanjilal {et~al.}(2021)Kanjilal, Dutta, \& Sharma}]{kanjilal2021growthcooling}
Kanjilal, V., Dutta, A., \& Sharma, P. 2021, Monthly Notices of the Royal Astronomical Society, 501, 1143, \dodoi{10.1093/mnras/staa3610}

\bibitem[{{Klein} {et~al.}(1994){Klein}, {McKee}, \& {Colella}}]{klein1994}
{Klein}, R.~I., {McKee}, C.~F., \& {Colella}, P. 1994, \apj, 420, 213, \dodoi{10.1086/173554}

\bibitem[{Li {et~al.}(2022)Li, Luo, Fossati, Sun, \& J{\'{a}}chym}]{Li2022TurbulenceGalaxy}
Li, Y., Luo, R., Fossati, M., Sun, M., \& J{\'{a}}chym, P. 2022, MNRAS, 000, 1

\bibitem[{{Li} {et~al.}(2020){Li}, {Hopkins}, {Squire}, \& {Hummels}}]{Li2020MNRAS.492.1841L}
{Li}, Z., {Hopkins}, P.~F., {Squire}, J., \& {Hummels}, C. 2020, \mnras, 492, 1841, \dodoi{10.1093/mnras/stz3567}

\bibitem[{McKee \& Ostriker(1977)}]{mckee1977asubstrate}
McKee, C.~F., \& Ostriker, J.~P. 1977, The Astrophysical Journal, 218, 148, \dodoi{10.1086/155667}

\bibitem[{P{\'{e}}roux \& Howk(2020)}]{peroux2020}
P{\'{e}}roux, C., \& Howk, J.~C. 2020, Annual Review of Astronomy and Astrophysics, 58, 363, \dodoi{10.1146/annurev-astro-021820-120014}

\bibitem[{{Pizzolato} \& {Soker}(2005)}]{AGNPizzolato2005}
{Pizzolato}, F., \& {Soker}, N. 2005, \apj, 632, 821, \dodoi{10.1086/444344}

\bibitem[{{Prasad} {et~al.}(2015){Prasad}, {Sharma}, \& {Babul}}]{Deovrat2015ApJ...811..108P}
{Prasad}, D., {Sharma}, P., \& {Babul}, A. 2015, \apj, 811, 108, \dodoi{10.1088/0004-637X/811/2/108}

\bibitem[{{Schmidt} {et~al.}(2006){Schmidt}, {Niemeyer}, \& {Hillebrandt}}]{schmidt2006}
{Schmidt}, W., {Niemeyer}, J.~C., \& {Hillebrandt}, W. 2006, \aap, 450, 265, \dodoi{10.1051/0004-6361:20053617}

\bibitem[{Sparre {et~al.}(2019)Sparre, Pfrommer, \& Vogelsberger}]{Sparre2019TheWind}
Sparre, M., Pfrommer, C., \& Vogelsberger, M. 2019, Monthly Notices of the Royal Astronomical Society, 482, 5401, \dodoi{10.1093/mnras/sty3063}

\bibitem[{Stone {et~al.}(2020)Stone, Tomida, White, \& Felker}]{stone2020}
Stone, J.~M., Tomida, K., White, C.~J., \& Felker, K.~G. 2020, The Astrophysical Journal Supplement Series, 249, 4, \dodoi{10.3847/1538-4365/ab929b}

\bibitem[{{Tan} {et~al.}(2023){Tan}, {Oh}, \& {Gronke}}]{tan2023infalling}
{Tan}, B., {Oh}, S.~P., \& {Gronke}, M. 2023, \mnras, 520, 2571, \dodoi{10.1093/mnras/stad236}

\bibitem[{{Townsend}(2009)}]{townsend2009}
{Townsend}, R.~H.~D. 2009, \apjs, 181, 391, \dodoi{10.1088/0067-0049/181/2/391}

\bibitem[{Tumlinson {et~al.}(2017)Tumlinson, Peeples, \& Werk}]{tumlinson2017themedium}
Tumlinson, J., Peeples, M.~S., \& Werk, J.~K. 2017, Annual Review of Astronomy and Astrophysics, AA, 1, \dodoi{10.1146/annurev-astro-091916-055240}

\bibitem[{{van der Velden}(2020)}]{Cmasher2020JOSS....5.2004V}
{van der Velden}, E. 2020, The Journal of Open Source Software, 5, 2004, \dodoi{10.21105/joss.02004}

\bibitem[{Veilleux {et~al.}(2005)Veilleux, Cecil, \& Bland-Hawthorn}]{veilleux2005galacticwinds}
Veilleux, S., Cecil, G., \& Bland-Hawthorn, J. 2005, Annual Review of Astronomy and Astrophysics, 43, 769, \dodoi{10.1146/annurev.astro.43.072103.150610}

\bibitem[{{Veilleux} {et~al.}(2020){Veilleux}, {Maiolino}, {Bolatto}, \& {Aalto}}]{veilleux2020}
{Veilleux}, S., {Maiolino}, R., {Bolatto}, A.~D., \& {Aalto}, S. 2020, \aapr, 28, 2, \dodoi{10.1007/s00159-019-0121-9}

\bibitem[{Vidal-Garc{\'{i}}a {et~al.}(2021)Vidal-Garc{\'{i}}a, Falgarone, Arrigoni Battaia, Godard, Ivison, Zwaan, Herrera, Frayer, Andreani, Li, \& Gavazzi}]{VidalGarcia2021}
Vidal-Garc{\'{i}}a, A., Falgarone, E., Arrigoni Battaia, F., {et~al.} 2021, Monthly Notices of the Royal Astronomical Society, 506, 2551, \dodoi{10.1093/mnras/stab1503}

\bibitem[{Virtanen {et~al.}(2020)Virtanen, Gommers, Oliphant, Haberland, Reddy, Cournapeau, Burovski, Peterson, Weckesser, Bright, van~der Walt, Brett, Wilson, Millman, Mayorov, Nelson, Jones, Kern, Larson, Carey, Polat, Feng, Moore, VanderPlas, Laxalde, Perktold, Cimrman, Henriksen, Quintero, Harris, Archibald, Ribeiro, Pedregosa, van Mulbregt, Vijaykumar, Bardelli, Rothberg, Hilboll, Kloeckner, Scopatz, Lee, Rokem, Woods, Fulton, Masson, H{\"{a}}ggstr{\"{o}}m, Fitzgerald, Nicholson, Hagen, Pasechnik, Olivetti, Martin, Wieser, Silva, Lenders, Wilhelm, Young, Price, Ingold, Allen, Lee, Audren, Probst, Dietrich, Silterra, Webber, Slavi{\v{c}}, Nothman, Buchner, Kulick, Sch{\"{o}}nberger, de~Miranda~Cardoso, Reimer, Harrington, Rodr{\'{i}}guez, Nunez-Iglesias, Kuczynski, Tritz, Thoma, Newville, K{\"{u}}mmerer, Bolingbroke, Tartre, Pak, Smith, Nowaczyk, Shebanov, Pavlyk, Brodtkorb, Lee, McGibbon, Feldbauer, Lewis, Tygier, Sievert, Vigna, Peterson, More, Pudlik, Oshima, Pingel, Robitaille, Spura, Jones, Cera,
  Leslie, Zito, Krauss, Upadhyay, Halchenko, V{\'{a}}zquez-Baeza, \& Contributors}]{virtanen2020scipypython}
Virtanen, P., Gommers, R., Oliphant, T.~E., {et~al.} 2020, Nature Methods, 17, 261, \dodoi{10.1038/s41592-019-0686-2}

\end{thebibliography}
\bibliographystyle{aasjournal}

 \end{document}